\documentclass[authoryear,preprint,10pt,review]{elsarticle}
\usepackage{amssymb,amsmath,gensymb,multirow, threeparttable}
\usepackage{lineno}

\makeatletter
\def\bbordermatrix#1{\begingroup \m@th
  \@tempdima 4.75\p@
  \setbox\z@\vbox{%
    \def\cr{\crcr\noalign{\kern2\p@\global\let\cr\endline}}%
    \ialign{$##$\hfil\kern2\p@\kern\@tempdima&\thinspace\hfil$##$\hfil
      &&\quad\hfil$##$\hfil\crcr
      \omit\strut\hfil\crcr\noalign{\kern-\baselineskip}%
      #1\crcr\omit\strut\cr}}%
  \setbox\tw@\vbox{\unvcopy\z@\global\setbox\@ne\lastbox}%
  \setbox\tw@\hbox{\unhbox\@ne\unskip\global\setbox\@ne\lastbox}%
  \setbox\tw@\hbox{$\kern\wd\@ne\kern-\@tempdima\left[\kern-\wd\@ne
    \global\setbox\@ne\vbox{\box\@ne\kern2\p@}%
    \vcenter{\kern-\ht\@ne\unvbox\z@\kern-\baselineskip}\,\right]$}%
  \null\;\vbox{\kern\ht\@ne\box\tw@}\endgroup}
\makeatother

\journal{: arXiv} 
\begin{document}
\begin{frontmatter}
\title{A simple and intuitive method to calculate $R_0$ in complex epidemic models}

\author[UFG]{Carlos Hernandez-Suarez\corref{cor1}}
 \ead{cmh1@cornell.edu}
 
 \author[UC]{Osval Montesinos L\'opez\corref{cor2}}
 \ead{oamontes1@ucol.mx}

\address[UFG]{Instituto de Innovaci\'on y Desarrollo, Universidad Francisco Gavidia, San Salvador, El Salvador}
\address[UC]{Facultad de Telem\'atica, Universidad de Colima, Colima 28040, Mexico}
  
 \cortext[cor1]{Corresponding author}

\begin{abstract}
Epidemic models are a valuable tool in the decision making process. Once a mathematical model for an epidemics has been established, the very next step is calculating a mathematical expression for the basic reproductive number, $R_0$, which is the average number of infections caused by an individual that is introduced in a population of susceptibles. Finding a mathematical expression for $R_0$ is important because it allows to analyze the effect of the different parameters in the model on $R_0$ so that we can act on them to keep $R_0 < 1$, so that the epidemic fades out. In this work we show how to calculate $R_0$ in complicated epidemic models by using only basic concepts of Markov chains.

\end{abstract}


\end{frontmatter}


\section{Introduction}
 
 This paper is not about \textit{building} epidemic models but about their analysis, that is, getting conclusions from them. In particular, this paper deals with calculating $R_0$, the most important quantity in an epidemic model. This paper is a significatively improved version of \cite{hernandez2002markov}.
 
 First we need to understand why $R_0$ is important in an epidemic model. The concept comes from the theory of branching process, in which we deal with the following problem: if an individual (particle or whatever) gives origin (birth) to a random number of descendants, and each of the descendants in turn gives birth to a random number of individuals and so on, what is the probability that the population will become extinct? First, observe that we mentioned a \textit{random} number of individuals, because if the number of descendants is a constant, then there is no problem at hand. We need to assume that every individual in every generation, leaves $X$ descendants before dying, and $X$ is a random variable with some probability distribution. We leave for appendix A1 the proof of the following result: if the expectation of the random variable $X$ is $\mu$, then, if $\mu \le 1$ the population will vanish eventually. A corollary is that if $\mu > 1$, then there is a chance that the population will prevail, and the chances increase with $\mu$.
 
The above model assumes that all individuals are equally capable to reproduce, that is, that the random variable $X$ follows the same distribution for all individuals, which is nat always true. Assume for instance that the resources to survive are limited, that is, suppose that we are talking about a bacteria population in a petri dish, where the population will start competing for resources (besides the toxicity of chemicals produced by bacteria) and where we know that even if $\mu>1$, the population of bacteria will perish because it can go beyond the petri dish. Of course, nobody wants to calculate what is the probability that the bacteria population will thrive and survive, because it is zero. But we may be interested in the conditions to a quick vanishing of the population, because perhaps we don't want it to grow. In this case, the average number of descendants would be tricky to calculate, because the first individual has on average more descendants that those from later generations, due to saturation. But there is a turnaround to the problem: if we find out that the first individual has an average number of descendants smaller or equal to one, then, there is nothing to worry about, because descendants at later generations will have even worst conditions to reproduce than the first one, and that is the reason why we only need to calculate the reproductive potential of the first one, because no other individual born at later generations can have a higher reproductive capability, when resources to reproduce becomes scarce.

This also applies to contagion: if the average number of infections is larger than one, the disease may become a large epidemics. Nevertheless, the first individual has always more chances to find susceptible individuals to infect, whereas subsequently infected individuals will find previously infected (and perhaps immune) individuals, so, it is harder for them to keep the same average number of infections than the first infected, but, if the average number of infections of the first individual, the one with more potential to produce more infections (when everybody else is susceptible) is smaller than one, then, all other subsequently infected, with less potential due to saturation, have an average number of infections less than one and no large epidemics will occur. 

In mathematical epidemiology the equivalent to the average number of descendants is thus the average number of infections. And the average number of descendants, $\mu$, becomes $R_0$ or \textit{basic reproductive number}  in mathematical epidemiology. Now it is clear why the definition of $R_0$ has been established in terms of the potential of the first individual \citep{MR1057044, heesterbeek1996concept}.

\subsection{The life circle}

There is some care we need to take in the calculation of the average number of individuals produced by an individual. Figure 1, is a typical example of the problem at hand, related to \textit{Eratyrus mucronatus}, the \textit{kissing bug} that transmits Chagas' disease \citep[see][]{hernandez2019building}. Individuals in the Adult stage produce individuals in the egg stage, but the average number of eggs produced is not $\mu$, because there is no guarantee that an egg will become an adult and thus, the average number of eggs produced by an adult is not a measure of the ability of an individual to replace itself. What we need to calculate is the average average number of eggs produced by an egg, or the average number of adults produced by an adult, or the average number of stage 1 individuals produced by a stage 1 individual and so on.

\begin{figure}[htbp]
\begin{center}
\includegraphics[width=12cm]{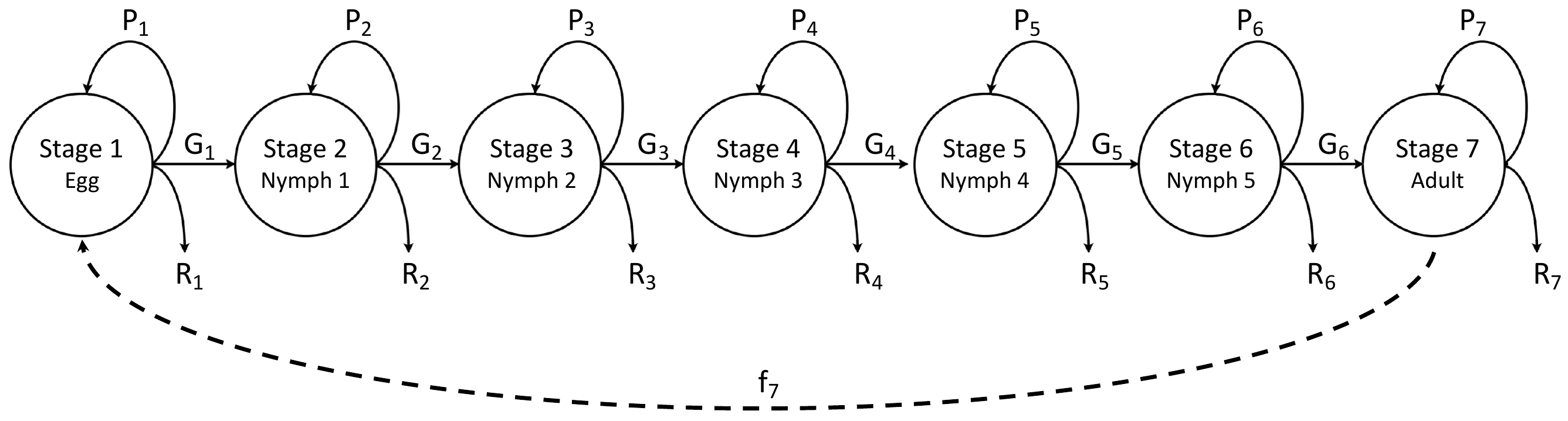}
\end{center}
\caption{The life cicle of \textit{Eratyrus mucronatus}. In this example, the Adult stage is the reproductive one, and all newborns are eggs. The dashed arrow indicates that all newborns belong to the first stage, and $f_7$ is the fertility rate.}\label{fig:01}
\end{figure}

The previous rationale applies to epidemic models, where  a particular individual moves along a series of states or compartments (susceptible, infected, isolated, vaccinated, dead, etc.) according to certain rules, and, for obvious reasons, we are interested in how many infected will be produced by an infected, and it has been clarified already that we need to analyze the infectious potential of the first infected only.

Define a \textit{contact} as any act between two individuals, that would cause the infectious of the susceptible if the contact involves an infectious and a susceptible. This could be sexual contact, sharing hands, sharing needles, talking, etc. With this in mind, observe that the number of contacts that an individual has per unit of time (day, hour, week, etc)  is a random variable $Y$ with expectation $\theta$, and this is the number that matters. Everything we need to know is the number of contacts that an infectious individuals has during the time s/he is infected, because, by assumptions in the model, $Y$ has the same distribution for every individual, regardless if it is the first infected or the one in the middle or the last one, because individuals have contacts whether they are infected or not, that is, contacts occur even before the epidemic starts. This means that to obtain $R_0$, we need to calculate the average number of contacts an individual has per unit of time and then multiply this number by the average time the individual spends infectious.

\section{Methodology}

We have highlighted the key to calculate $R_0$, regardless how complicated the model is: if $a$ is the rate (number of events per unit of time) at which an individual has contacts, and $w$ is the average time an infected individual is infectious, then $R_0 = a \ w$. The problem is not $a$, since most of the times it is one of the parameters in the model. The problem is $w$, the average time infectious, since the model may be very complex and it is not clear how to calculate it. For that, we need some theory of Markov chains. Nevertheless, in many cases, $R_0$ can be calculated by inspection.  For instance, consider the classical SIR model in Figure 1. Above each arrow between any two compartments, there is the rate at which individuals move from one compartment to another. The word \textit{rate} needs to be clearly defined, and we use for this \textit{rate} in the sense of a stochastic model: if an individual leaves a box at a rate $\alpha$, it means that the average time an individual spends in that compartment is exponential with mean $1/\alpha$. 

\begin{figure}[htbp]
\begin{center}
\includegraphics[width=9cm]{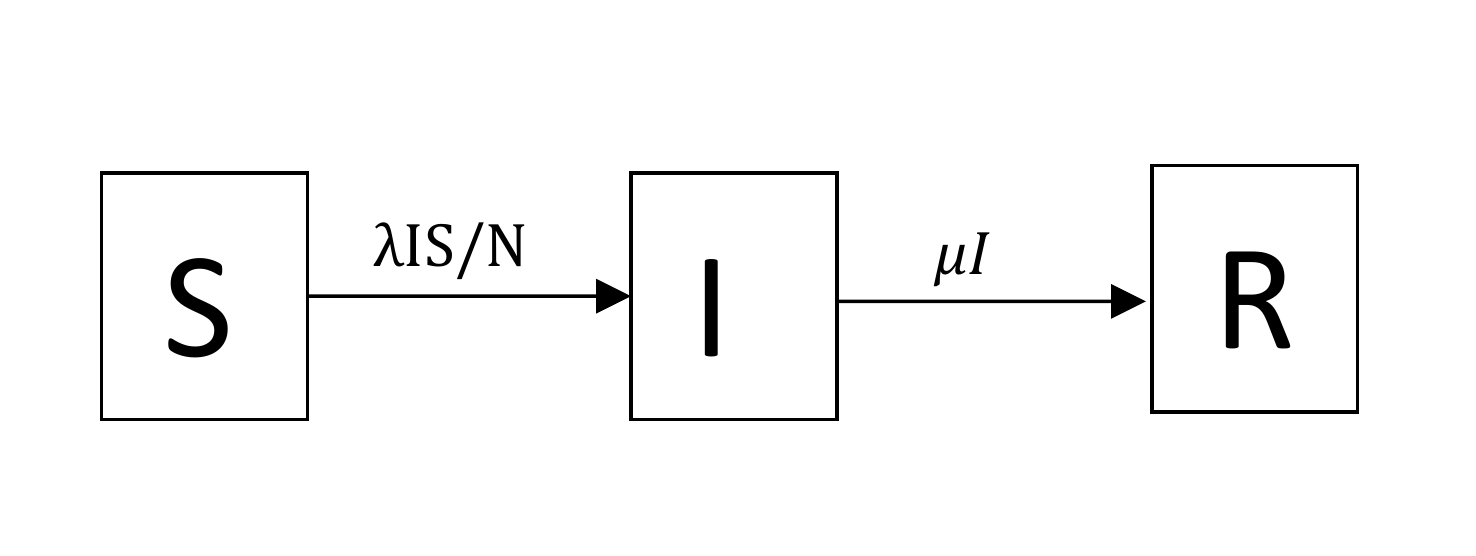}
\end{center}
\caption{The Susceptible- Infected -Removed epidemic model. $R_0 = \lambda/\mu$.}\label{fig:02}
\end{figure}

Now, in the SIR model, what is the average time an individual lasts infectious? We can see that the individual exit rate from compartment I is $\mu$, so on average, the individual remains infected for an average time $1/\mu$. But according to the model in Figure 2, every individual has contacts with another at a rate $\lambda$, therefore, the average number of contacts of an infectious individual is $\lambda/\mu$, which is $R_0$. Applying the same rationale, word by word, the $R_0$ value is the same for the SEIR model of Figure 2, where E stands for an infected, non infectious (``latent'') stage.

\begin{figure}[htbp]
\begin{center}
\includegraphics[width=9cm]{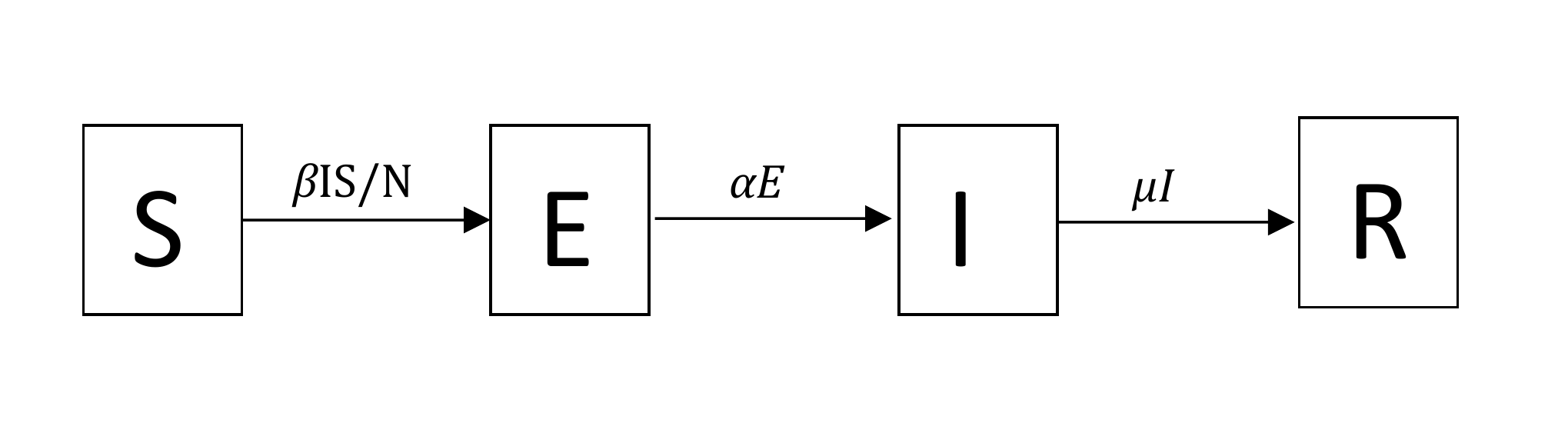}
\end{center}
\caption{The Susceptible- Latent -Infected -Removed epidemic model. $R_0$ is also $\lambda/\mu$.}\label{fig:03}
\end{figure}

Now, consider a particular case of a modified SEIR model depicted in Figure 4, where an individual in the latent stage can be detected and removed before going to the I state with probability $p$. This is a big change: in the SIR model, an infectious (I) individual produces infected individuals, whereas in the SEIR model, an infected individual produces latent (E) individuals. Since in an SEIR model every latent individual will become type I eventually with probability one (nothing in the model indicates otherwise) there is actually no difference in $R_0$ between SIR and SEIR, but in the model of Figure 4, an infectious individual produces latent individuals and now there is a chance that they may not become infectious. This is a similar problem to the ``life cycle'' problem mentioned in the previous section. Since what is needed is the average of infectious produced by an infectious, in the modified  SEIR the chance of being removed before becoming infectious, must be considered. Thus, the average amount of time that a recently infected individual spends in the infectious compartment I changes from $1/\mu$ to $(1-p)/\mu$. Since the number of contacts is still $\lambda$, the $R_0$ for this modified SEIR is $\lambda (1-p)/\mu$.

\begin{figure}[htbp]
\begin{center}
\includegraphics[width=9cm]{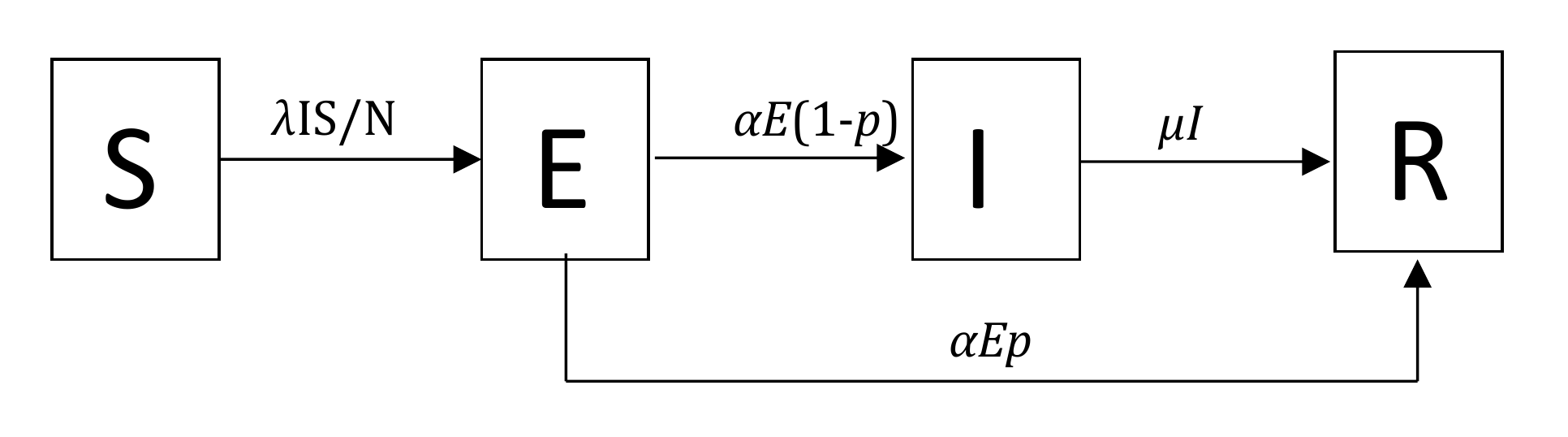}
\end{center}
\caption{The Susceptible- Latent -Infected -Removed epidemic model with detection and removal of latent. Now there is a chance that a newly infected will not become infectious. $R_0$ for this model is $\lambda (1-p)/\mu$.}\label{fig:04}
\end{figure}

Another example, a bit more complicated is the model of Figure 5, in which an individual infected may alternate between infectious (I) and non-infectious (C) states several times before being removed. That is, the individual may move through the states like S-I-C-I-C-I-R, or S-I-C-I-C-I-C-I-C-I-R or S-I-R, just to illustrate the problem. Once the expected number of times the individual visits stage I, say $k$, we multiply this value by $(\mu+\alpha)^{-1}$ which is  the average time in each visit to I, to give the average time spent in the infectious stage, and then, this the result is multiplied by the contact rate to yield $R_0$.  

It requires some training and experience to calculate $R_0$ just  by inspection, that is why we need the theory of Markov chains. The resulting theory can be applied to any model, simple or not.

\begin{figure}[htbp]
\begin{center}
\includegraphics[width=9cm]{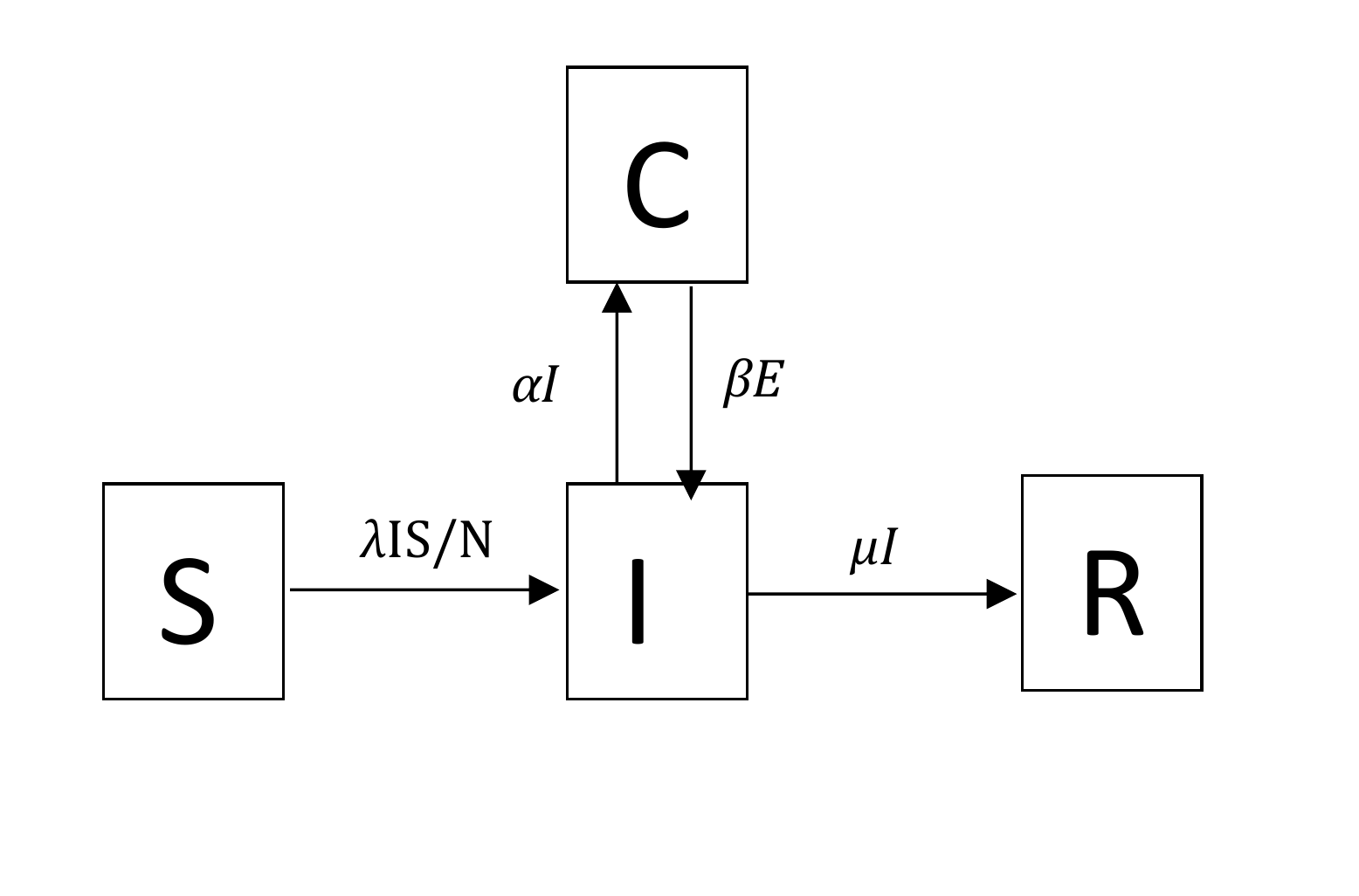}
\end{center}
\caption{A model in which an infected individual switches between a carrier (C), non-infectious stage and an infectious stage (I). The number of visits to state I is a geometric random variable.}\label{fig:05}
\end{figure}

\subsection{Markov chains}

We will focus on discrete time, discrete space Markov chains. An excellent review of these process at the level required in this paper is found in \cite{ross-07, grinstead1997introduction} and a deeper treatment in \cite{iosifescu1980finite}. Imagine an individual moving across a series of states (see Figure 6) in some random fashion: when it is in stage $i$, it moves to stage $j$ with some constant probability $p_{ij}$. The time spent in each stage is, by now, irrelevant. For now, assume that the time spent on each stage is unitary. This is a very useful model with applications in a wide range of areas. Observe that, regardless on where the individual starts, the process never ends, the individual keeps moving: 3,4,2,3,2,3,4,2,1,4,2,3,...  There are many interesting questions related to this model, for instance, what is the probability that an individual starting in stage 3 will be on stage 1 after $k$ transitions? Or, what will the the proportion of time spent on each stage when the number of transitions goes to infinity? \begin{figure}[htbp]
\begin{center}
\includegraphics[width=8cm]{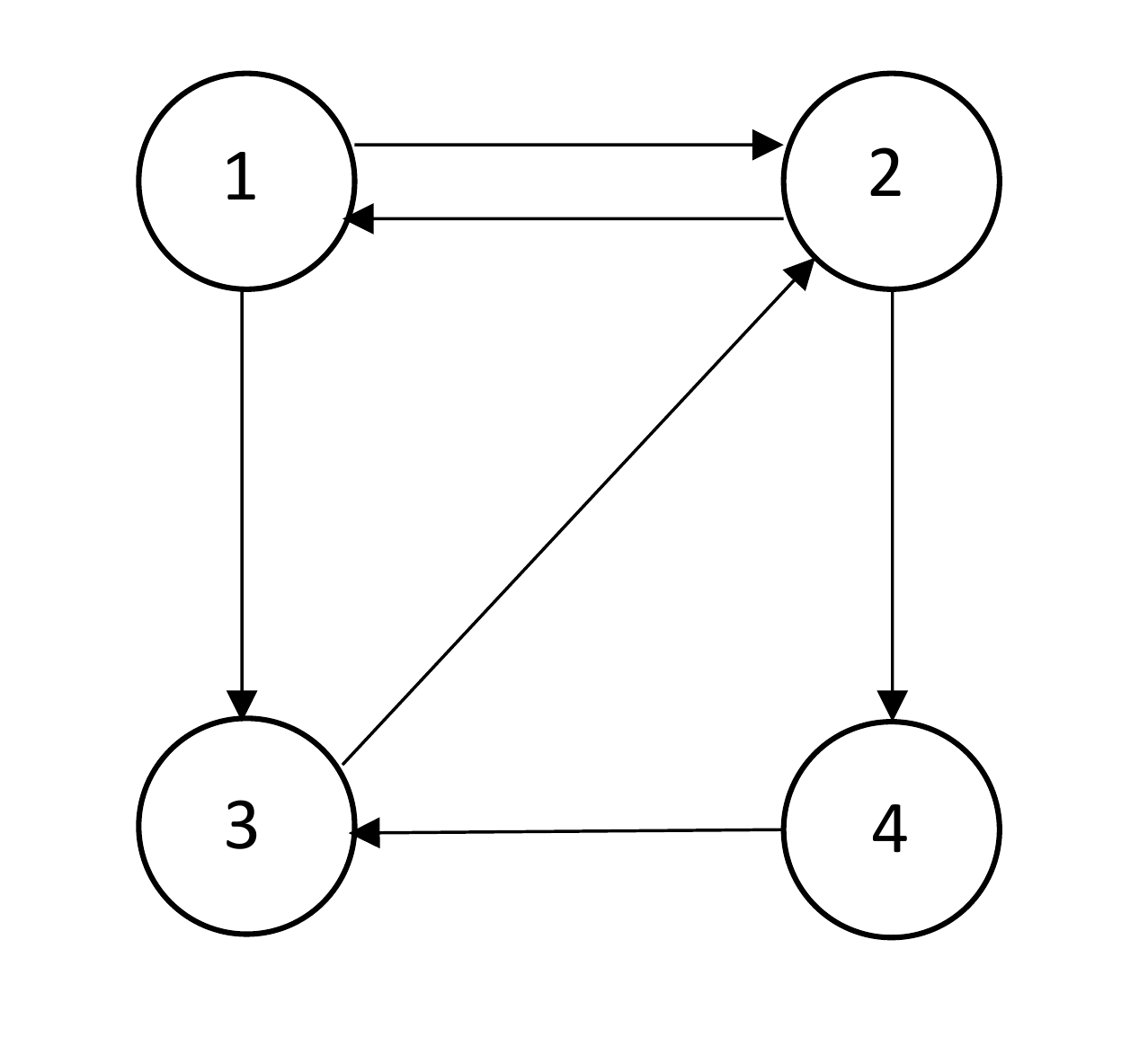}
\end{center}
\caption{A Markov chain with four states.}\label{fig:06}
\end{figure}

But there is a class of Markov chains we will focus on: those that contain one or more absorbing states, where an absorbing state is a black hole that swallows the individual and this cannot leave this stage, and thus the process stops (see Figure 7). These Markov chains also raise many questions as, for instance: what is the probability that the individual ends trapped on some state $a$ rather than in state $b$ (assuming that states $a$ and $b$ are absorbing states), or, what is the average number of visits made to some state $x$ before the process stops? The answers to these questions are simple and straightforward, although we refer to other texts for the proof.

\begin{figure}[htbp]
\begin{center}
\includegraphics[width=8cm]{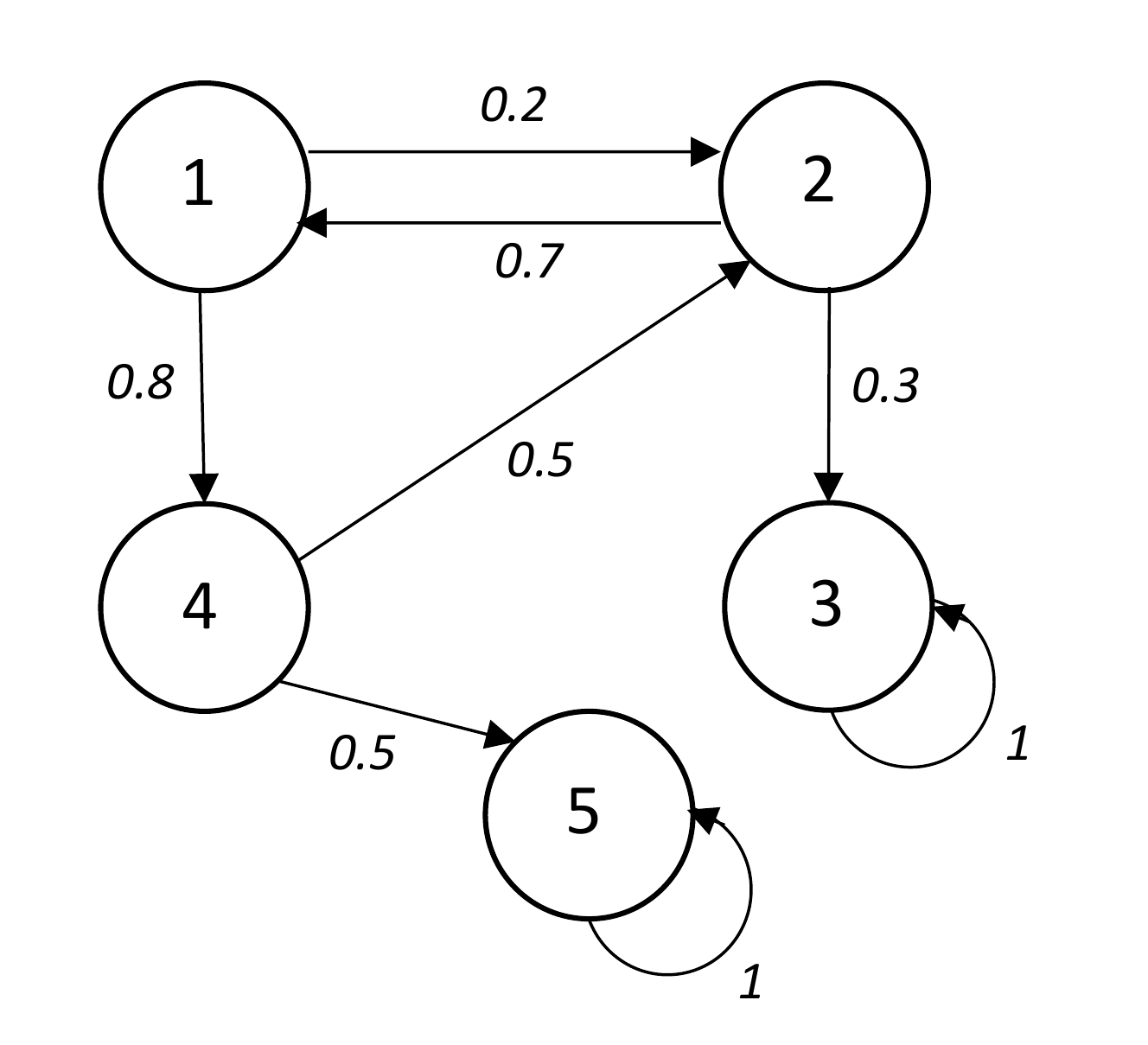}
\end{center}
\caption{A Markov chain with five states, where states 3 and 5 are \textit{absorbing} states. States 1,2 and 4 are called \textit{transient} states.}\label{fig:07}
\end{figure}

Consider for instance the Markov chain of Figure 7. We first need to construct a matrix of transitions between states, where $p_{ij}$ is the probability that once in state $i$ the individual moves to state $j$.

 $$
\mathbf{P}= \bbordermatrix{ & 1 & 2 & 3 & 4 & 5 \cr
1 & 0 & 0.2 & 0 & 0.8 & 0 \cr
2 & 0.7 & 0 & 0.3 & 0 & 0 \cr
3 & 0 & 0 & 1 & 0 & 0 \cr
4 & 0 & 0.5 & 0 & 0& 0.5 \cr
5 & 0 & 0 & 0 & 0 & 1 \cr}
$$

\textit{Absorbing} states are easily located and correspond to the 1's in the diagonal (states 3 and 5). The rest of the states is called \textit{transient} states. We first need to rearrange rows and columns of the matrix $\mathbf{P}$ so that the absorbing states are located at the bottom right corner, to the form:

 $$
\mathbf{P}= \bbordermatrix{ &  &  \cr
               & \mathbf{U} & \mathbf{R}\cr
                & \mathbf{Z} & \mathbf{I} \cr}
$$
\noindent
where $\mathbf{U}$ is a $k \times k$ matrix containing the transitions between the $k$ transient states, $\mathbf{R}$ is a $k \times r$ matrix containing the transitions between the $k$ transient states to the $r$ absorbing states, $\mathbf{Z}$ is a $r \times k$ matrix of zeros and  $\mathbf{I}$ is the identity matrix of size $r$.  Rearranging matrix $\mathbf{P}$ yields:

 $$
\mathbf{P}= \bbordermatrix{ & 1 & 2 & 4 & & 3 & 5 \cr
1 & 0 & 0.2 & 0 & | & 0.8 & 0 \cr
2 & 0.7 & 0 & 0.3 & | & 0 & 0 \cr
4 & 0 & 0.5 & 0 & | & 0& 0.5 \cr
 & - & - & - & -& -& - \cr
3 & 0 & 0 & 0 & |& 1 & 0 \cr
5 & 0 & 0 & 0 &| & 0 & 1 \cr}
$$

Then: 

 $$
\mathbf{U}= \bbordermatrix{ &  &  \cr
          &  0    & 0.2 & 0\cr
          &  0.7    & 0 & 0.3\cr
          &  0    & 0.5 & 0\cr } ,\ \ \ \
          \mathbf{R}= \bbordermatrix{ &  &  \cr
          &  0.8    & 0\cr
          &  0    & 0\cr
           &  0    & 0.5\cr}
$$
 
 \subsubsection{The fundamental matrix}
 
 Among the many useful results we can derive from the previous decomposition of matrix $\mathbf{P}$, the one we need is the following result: 
 \\
 \\
 \textit{The average number of visits to each transient state before absorption, is given by} $\mathbf{N}=( \mathbf{I}- \mathbf{U})^{-1}$. 
  \\
  \\
 The proof is in Appendix A2. Matrix $\mathbf{N}$ is called the \textit{fundamental matrix} of $\mathbf{P}$.
 
 $$
\mathbf{N}= \bbordermatrix{ & 1 & 2 &4 \cr
         1 & 1.197 &   0.282 &   0.084\cr
          2&   0.986   & 1.408  &  0.422\cr
          4&  0.4930 &  0.704  &  1.211\cr }
$$
\noindent
which is read as follows: the average number of visits to state 2, if the initial state is state 1, is 0.282 If the initial state is 4, the average number of visits to stage 4 is 1.211

\subsubsection{From rates to probabilities}

A useful result from probability theory, in particular, from properties of exponential distributions is related to how likely is that an individual moves from one state to another, given that we do not know transition probabilities, but only exit rates. Figure 8 illustrate the problem.  In this Figure, An individual moves from compartment A to B or C at rates $\alpha$ and $\beta$ respectively. But it can only move to one of them. The result is that the probability that it moves to A is $\alpha/(\alpha+\beta)$ and to B with the remaining probability $\beta/(\alpha+\beta)$. This result can be extended to any number of compartments.

\subsubsection{Average time in a compartment}

The last result we need is related to the average time in a compartment. Figure 8 is useful to illustrate the result: the average time in a compartment is equal to the inverse of the sum of all exit rates, regardless the number of compartments. In this case is $(\alpha+\beta)^{-1}$.

\begin{figure}[htbp]
\begin{center}
\includegraphics[width=5cm]{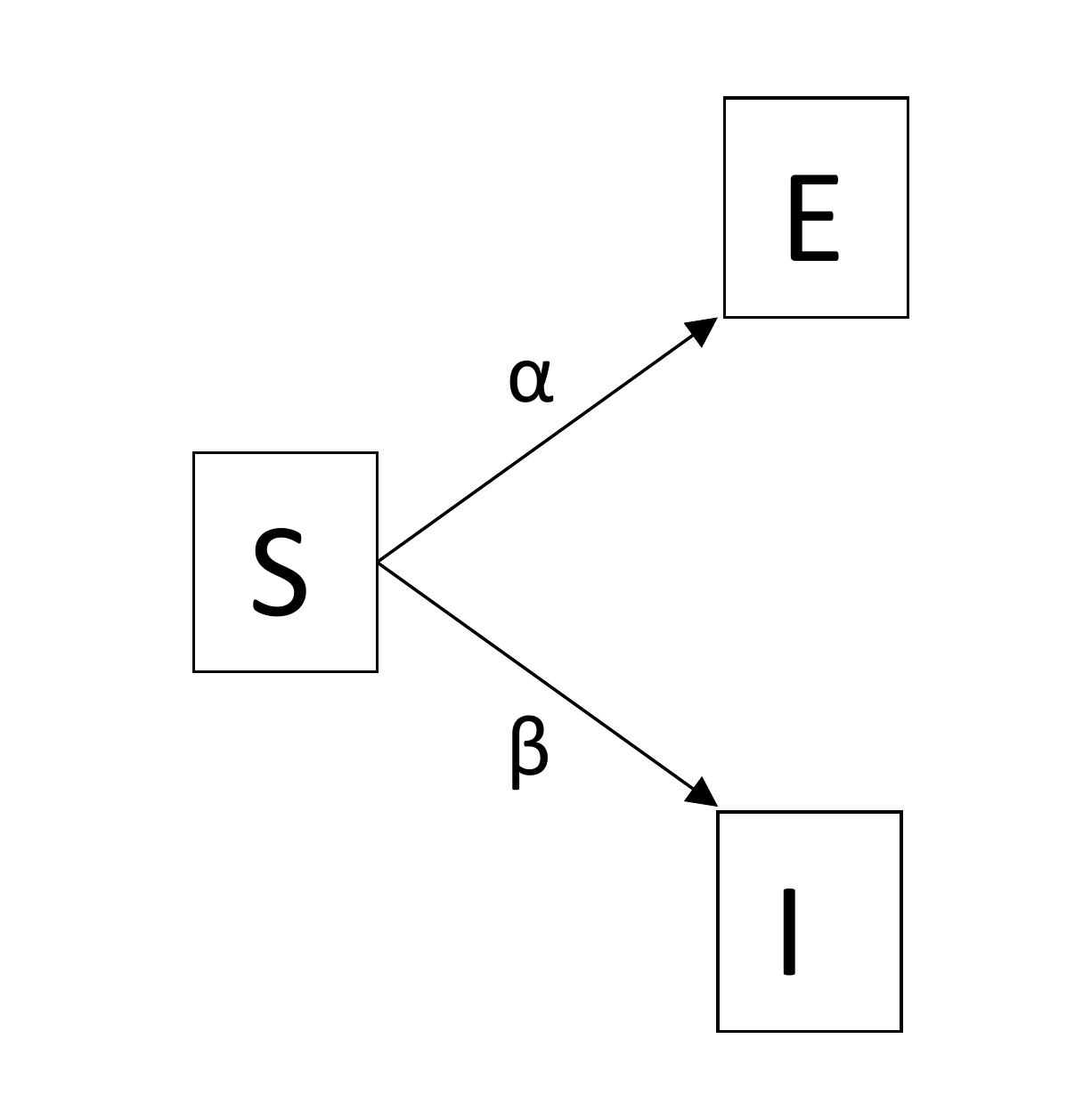}
\end{center}
\caption{An individual moves from compartment A to B or C at rates $\alpha$ and $\beta$ respectively. The probability that it moves to A is $\alpha/(\alpha+\beta)$. The average time in compartment A is $(\alpha+\beta)^{-1}$ }\label{fig:08}
\end{figure}

We are now in possession of the three elements that will be used to calculate $R_0$. Once a model has been established, we need to follow four steps:

 \begin{enumerate}[(i)]
   
 \item
 Convert all rates in the model to probabilities, to build a transition matrix $\mathbf{P}$ between states.
 \item
 Calculate the average number of visits to the infectious state using the fundamental matrix $\mathbf{N}$.
 \item
Calculate the average time spent in a single visit to the infectious state and use the result in (ii) to obtain the average time total expected time in the infectious state.
\item
Multiply the result in (iii) by the contact rate, to obtain the total number of contacts of an infectious individual. This is $R_0$.
 
\end{enumerate}

\section{Examples}

\subsection{Modified SEIR}

The first example is the modified SEIR (Figure \ref{fig:04}). As previously shown, it does not require the use of Markov chains because is simple enough to be calculated by inspection, but it will be used with the purpose of preparation of more complicated examples.

Step (\textit{i}) is the calculation of the transition matrix $\mathbf{P}$ in this case is:

 $$
\mathbf{P}= \bbordermatrix{ & S & E & I & & R  \cr
S & 0 & 1 & 0 & | & 0  \cr
E & 0 & 0 & 1-p & | & p \cr
I & 0 & 0 & 0 & | & 1 \cr
 & - & - & - & -& - \cr
R & 0 & 0 & 0 & |& 1  \cr}
$$
where:
 $$
\mathbf{U}= \bbordermatrix{ & S  &E & I  \cr
        S  &  0    & 1 & 0\cr
        E  &  0    & 0 & 1-p\cr
         I &  0    & 0 & 0\cr }
$$

To calculate step (\textit{ii}) we need the \textit{fundamental matrix} $\mathbf{N}$ which is:

 $$
\mathbf{U}= \bbordermatrix{ & S  &E & I  \cr
        S  &  1    & 1 & 1-p\cr
        E  &  0    & 1 & 1-p\cr
         I &  0    & 0 & 1\cr }
$$
\noindent
where we can see the expected number of visits to the infectious state I starting from state S (we always start in this state) is $1-p$.

For step (\textit{iii}), the average time spent in a single visit to the infectious state, this is the inverse of the sum of all exit rates from I, which is $1/\mu$ (we need to consider only the \textit{individual} exit rate, since the total exit rate is $\mu I$, the result follows). Multiplying this by the average number of visits to the infectious state gives $(1-p)/\mu$ as the average total amount of time that an individual stays infectious.

$R_0$ from step (\textit{iv}) is just the product of the contact rate and the previous result, $\lambda (1-p)/\mu$.

\subsection{Modified SEIR with natural mortality}

The second example is another version of the SEIR in which individuals in states E or I may die (Figure 9). We can actually start directly writing matrix $\mathbf{U}$ the matrix that excludes the absorbing states (in this case state R and the death states). Observe that the death state is not indicated in the figure but is where all the arrows pointing down are directed to. The remaining states (transient) are S, E and I. Matrix $\mathbf{U}$ is:

\begin{figure}[htbp]
\begin{center}
\includegraphics[width=9cm]{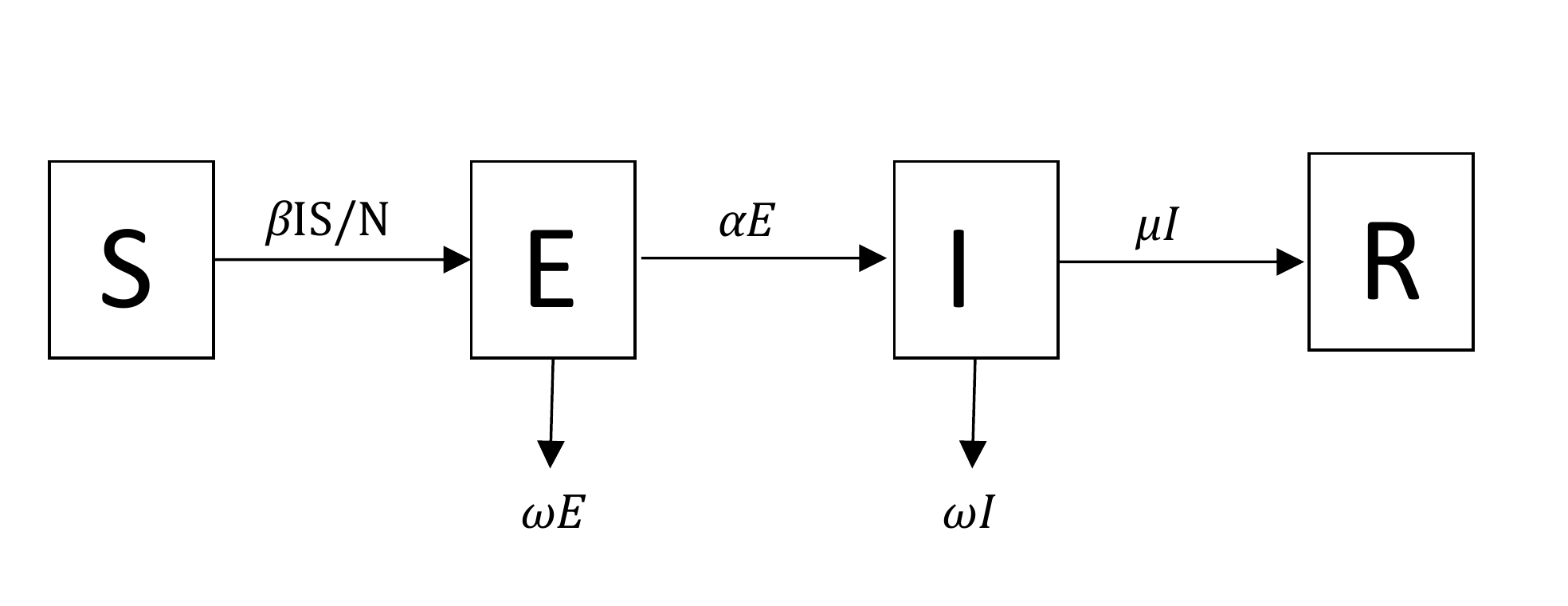}
\end{center}
\caption{An SEIR model with disease induced mortality. In this model $N=S+E+I$ }\label{fig:09}
\end{figure}

 $$
\mathbf{U}= \bbordermatrix{ & S & E & I \cr
		S & 0 & 1 & 0\cr
               E& 0 & 0&\frac{\alpha}{\alpha+\omega}\cr
                I& 0 & 0  &0\cr}
$$
The reason why the probability that the individual moves from state E to I is $\alpha/(\alpha+\omega)$ has been explained in Figure 8.

The fundamental matrix $\mathbf{N}$ is:

 $$
\mathbf{U}= \bbordermatrix{ & S & E & I \cr
		S & 1 & 1 & \frac{\alpha}{\alpha+\omega}\cr
		E & 0 & 1 & \frac{\alpha}{\alpha+\omega}\cr
	          I& 0 & 0  &1\cr}
$$
\noindent
where we can see the expected number of visits to the infectious state I starting in state S is $\alpha/(\alpha+\omega)$. Observe that the average time per visit to state I is $(\mu+\omega)^{-1}$, therefore, the average total time spent in state I is

$$
 \frac{\alpha}{(\alpha+\omega)(\mu+\omega)},
$$

multiplying this by the contact rate $\beta$, yields
$$
R_0= \frac{\beta \alpha}{(\alpha+\omega)(\mu+\omega)}.
$$

\subsection{An Ebola transmission model}

We will analyze another example where individuals are infectious at different stages and with a different degree of infectiousness. This is an Ebola model in infectious individuals infect susceptible ones while alive outside hospitals, in hospitals and dead but not buried \citep{legrand2007understanding}. The infectious states are $I$, $H$ and $F$. State $E$ is included since it communicates with one of the infectious states. The final set is $\{E,I,H,F\}$ (see Figure 10).

 \begin{figure}[htbp]
\begin{center}
\includegraphics[width=11cm]{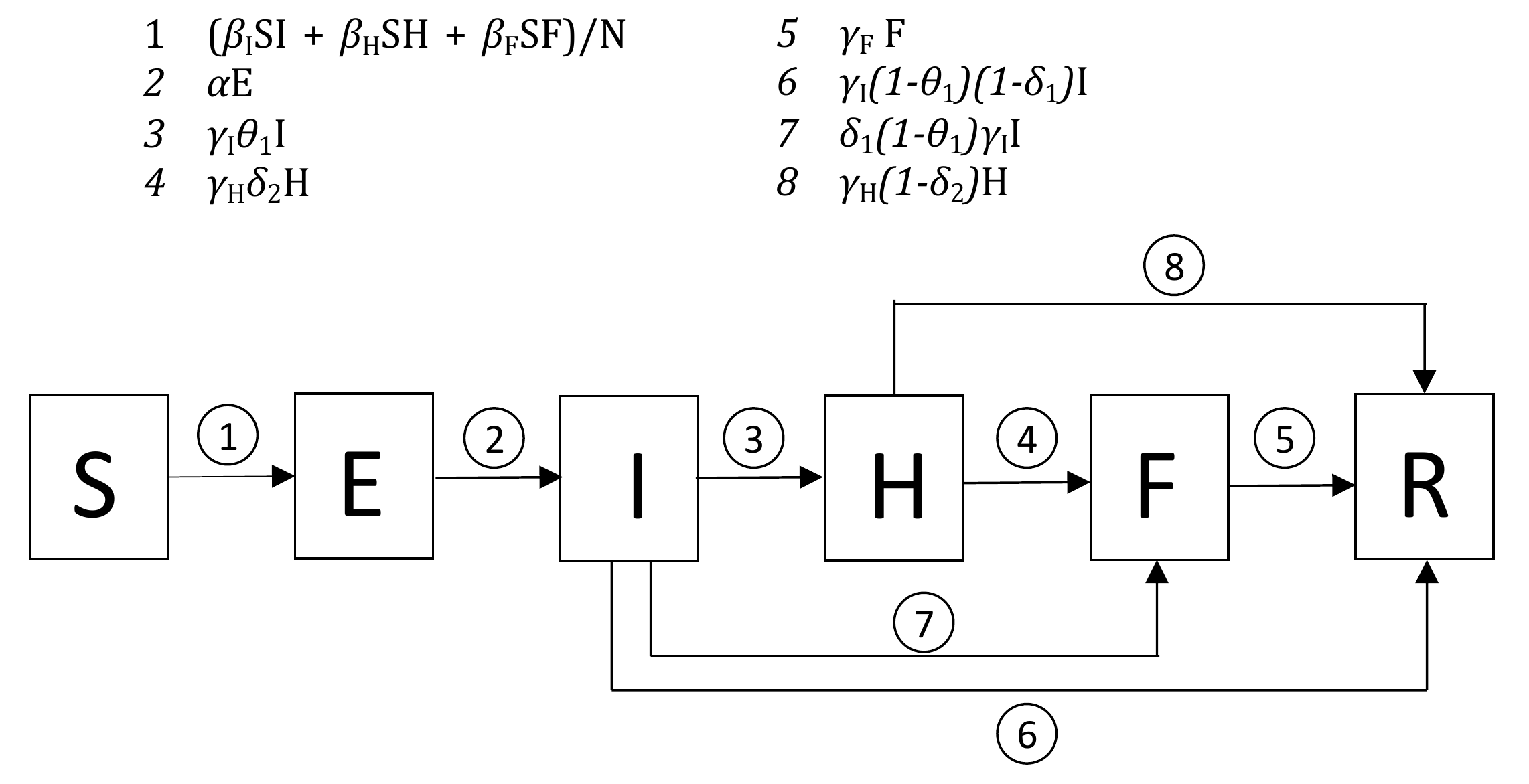}
\end{center}
\caption{An Ebola transmission model in which hospitalized and unburied are a source of infection. $S$, Susceptible individuals; $E$, Exposed individuals; $I$, Infectious; $H$,
hospitalized; $F$, dead but not yet buried; $R$, removed. The parameters are: $\beta_I$, transmission coefficient in the community; $\beta_H$, transmission coefficient at the hospital; $\beta_F$,
transmission coefficient during funerals. $\theta_1$ fraction of infectious cases that are hospitalized. $\delta_1, \delta_2$ are computed in order that the overall case-fatality ratio is $\delta$. The incubation rate is $\alpha$. The mean duration from symptom onset to hospitalization is  $\gamma_h^{-1}$, $\gamma_{dh}^{-1}$ is the mean duration from hospitalization to death, and $\gamma_i^{-1}$ denotes the mean duration of the infectious period for survivors. The mean duration from hospitalization to end of infectiousness for survivors is $\gamma_{ih}^{-1}$ and $\gamma_f^{-1}$  is the mean duration from death to burial. Transmission coefficients are expressed in weeks$^{-1}$ \citep{legrand2007understanding}.}\label{fig:10}
\end{figure}

We can skip directly to the transition matrix $\mathbf{U}$, that excludes the absorbing states. Thus:

 $$
\mathbf{U}= \bbordermatrix{ & S& E & I  & H &F \cr
	S& 0 & 1 & 0 & 0 & 0\cr
               E&0 & 0 & 1 & 0 & 0 \cr
                I&  0 & 0 & 0 & \theta _1 & \delta _1 \left(1-\theta _1\right)\cr
                 H& 0& 0 & 0 & 0 & \delta _2\cr
               F& 0& 0&0 &0 &0\cr}
$$
The fundamental matrix $\mathbf{N}$ is:

$$
\mathbf{N}= \bbordermatrix{ & S& E & I  & H &F \cr
S& 1 & 1 & 1 & \theta _1 & \delta _2 \theta _1+\delta _1 \left(1-\theta _1\right) \cr
E&  0 & 1 & 1 & \theta _1 & \delta _2 \theta _1+\delta _1 \left(1-\theta _1\right) \cr
 I& 0 & 0 & 1 & \theta _1 & \delta _2 \theta _1+\delta _1 \left(1-\theta _1\right) \cr
H&  0 & 0 & 0 & 1 & \delta _2 \cr
 F& 0 & 0 & 0 & 0 & 1 \cr}
$$
\\
The infectious states are I, H and F.  From the first row of $\mathbf{N}$ we can obtain the expected number of visits to each one of these states, that are respectively: $1,\theta _1$ and $ \delta _2 \theta _1+\delta _1 (1-\theta _1)$. Since the expected times on each visit to these states are, respectively, $1/\gamma_I,1/\gamma_H$ and $1/\gamma_F$, we finally arrive to the total expected time spent in each one of the infectious states I, H and F. These are respectively:

$$
\big[1/\gamma_I, \ \theta_1/\gamma_H, \ (\delta _2 \theta _1+\delta _1 (1-\theta _1))/\gamma_F\big].
$$
The final expression for $R_0$ is obtained by multiplying each of those terms by their respective contact rate and adding them. Each of the terms is the expected number of infections produced by an individual in each state, that is, its contribution to $R_0$:

$$
R_0 = \beta_I/\gamma_I+\beta_H \theta_1/\gamma_H+ \beta_F(\delta _2 \theta _1+\delta _1 (1-\theta _1))/\gamma_F).
$$

\subsection{A model for COVID-19 transmission}

Another example is a model for Covid-19 transmission. This model was chosen because it seems complex enough, with eight states and twelve possible transitions \citep{ndairou2020mathematical}. The model has the following states: susceptible class (S), exposed class (E), symptomatic and infectious class (I), super-spreaders class (P), infectious but asymptomatic class (A), hospitalized (H), recovery class (R), and fatality class (F). Absorbing states are A, R and F, leaving the transient states S, E, I, H and P. Observe the transition probabilities between the transient states: $P(S\rightarrow E)=1,P(E\rightarrow I)=p_1,P(E\rightarrow P)=p_2,P(I\rightarrow H)=\gamma_a/(\gamma_a+\gamma_i+\delta_i), P(P \rightarrow H)=\gamma_a/(\gamma_a+\gamma_i).$ We can skip directly to writing the matrix $\mathbf{U}$:
 
  \begin{figure}[htbp]
\begin{center}
\includegraphics[width=13cm]{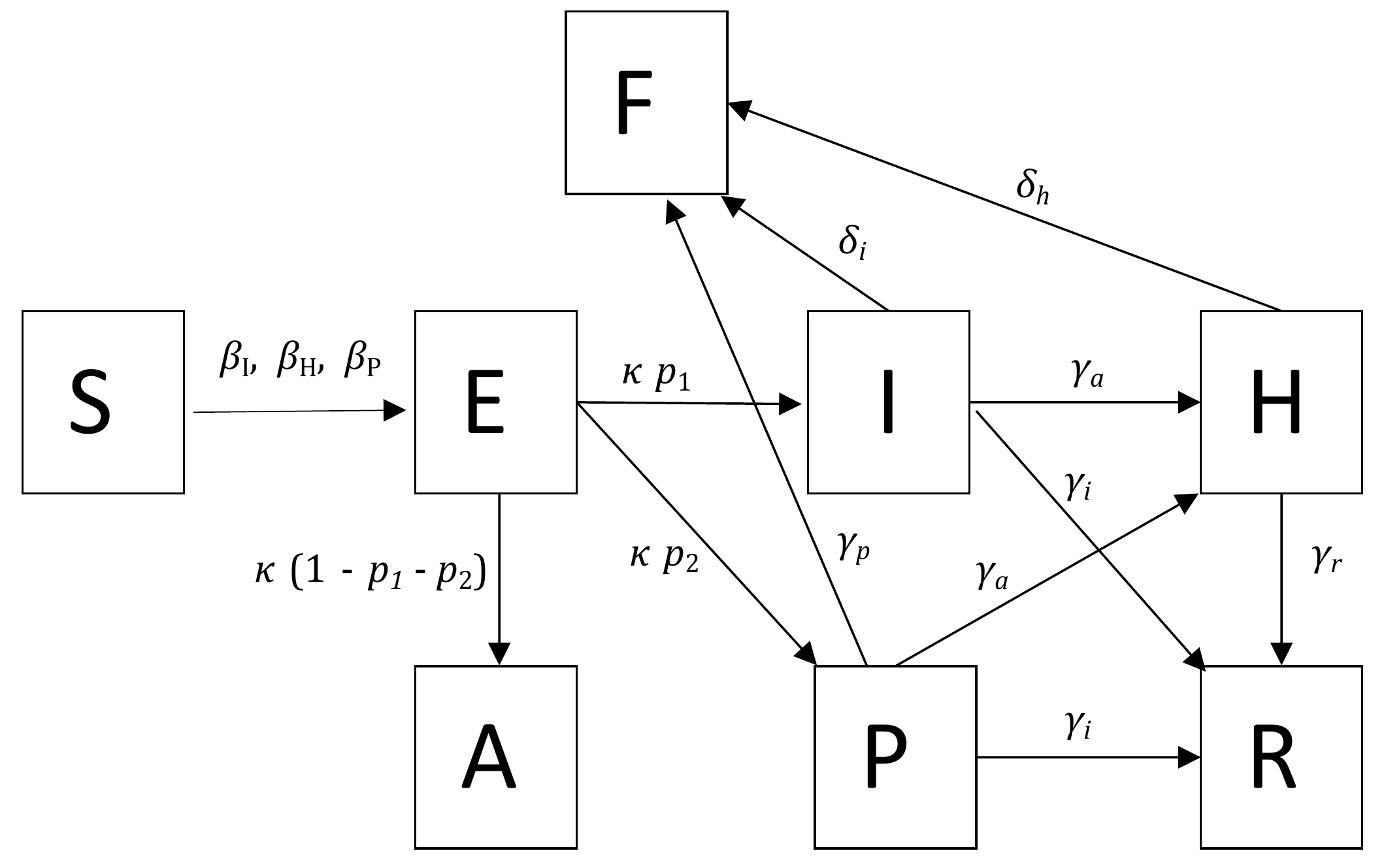}
\end{center}
\caption{A Covid-19 mode. Susceptible class (S), exposed class (E), symptomatic and infectious class (I), super-spreaders class (P), infectious but asymptomatic class (A), hospitalized (H), recovery class (R), and fatality class (F).  $\beta_I, \beta_H$ and $\beta_P$ are the contact rates of individuals in each of those states. }\label{fig:11}
\end{figure}
 
 $$
\mathbf{U}= \bbordermatrix{ & S& E & I  & H &P \cr
	S& 0 & 1 & 0 & 0 & 0\cr
               E&0 & 0 & p_1 & 0 & p_2 \cr
                I&  0 & 0 & 0 & \frac{\gamma_a}{\gamma_a+\gamma_i+\delta_i} & 0\cr
                 H& 0& 0 & 0 & 0 & 0\cr
               P& 0& 0&0 &\frac{\gamma_a}{\gamma_a+\gamma_i} &0\cr}
$$
 and the first row of fundamental matrix $\mathbf{N}$ is:

$$
\bigg[1,\ 1,\ p_1, \ \gamma_a\ p_1/(\delta_i + \gamma_a + \gamma_i) +\gamma_a \ p_2/(\gamma_a + \gamma_i)], \ p_2 \bigg]
$$
\noindent
which correspond to the expected number of visits to each of the transient states S, E, I, H and P, starting in state S.  The last three elements are of interest since they correspond the the infectious states I,H and P:

$$
\mathbf{v}'=\bigg[ p_1, \ \gamma_a\ p_1/(\delta_i + \gamma_a + \gamma_i) +\gamma_a \ p_2/(\gamma_a + \gamma_i)], \ p_2 \bigg]
$$ 
On the other hand, the average time per visit to each of the infectious stages I, H and P is, respectively:

$$
\mathbf{z}'=\bigg[(\delta_i + \gamma_a + \gamma_i)^{-1}, \  (\delta_h + \gamma_r )^{-1}, \   (\gamma_p + \gamma_a + \gamma_i)^{-1}    	\bigg]
$$ 
\noindent
and the average time spent on each of the infectious stages I,H and P is the Kronecker or element-by-element multiplication:

$$
\mathbf{w} = \mathbf{v} \otimes \mathbf{z}
$$

 Since the contact rate of the infectious states are respectively $\beta_I, \beta_H$ and $\beta_P$, we can make:
 
 $$
 \boldsymbol{\beta}' = [\beta_I \  \beta_H \ \beta_P]
 $$
 
 and then finally we get to $R_0$:
 
 $$
 R_0 = \boldsymbol{\beta}' \mathbf{w}
 $$
 
 \subsection{Tuberculosis transmission model}
 
 This example is a model for Tuberculosis  \cite{blower1995intrinsic}, and it is interesting because the infectious individuals can recover temporarily and become infectious again, alternating between the two states (see Figure 11). In all states there is natural mortality and in some there is additional disease- induced mortality. The states are : S, susceptible; L, latent; $T_i$, infected and infectious; $T_n$, infected but non-infectious; $R_i$, infectious individuals that become recovered and non-infectious for a while; $R_n$, non-infectious individuals that become recovered for a while. All these states are transient, and the transition matrix $\mathbf{U}$ is:

  \begin{figure}[htbp]
\begin{center}
\includegraphics[width=10cm]{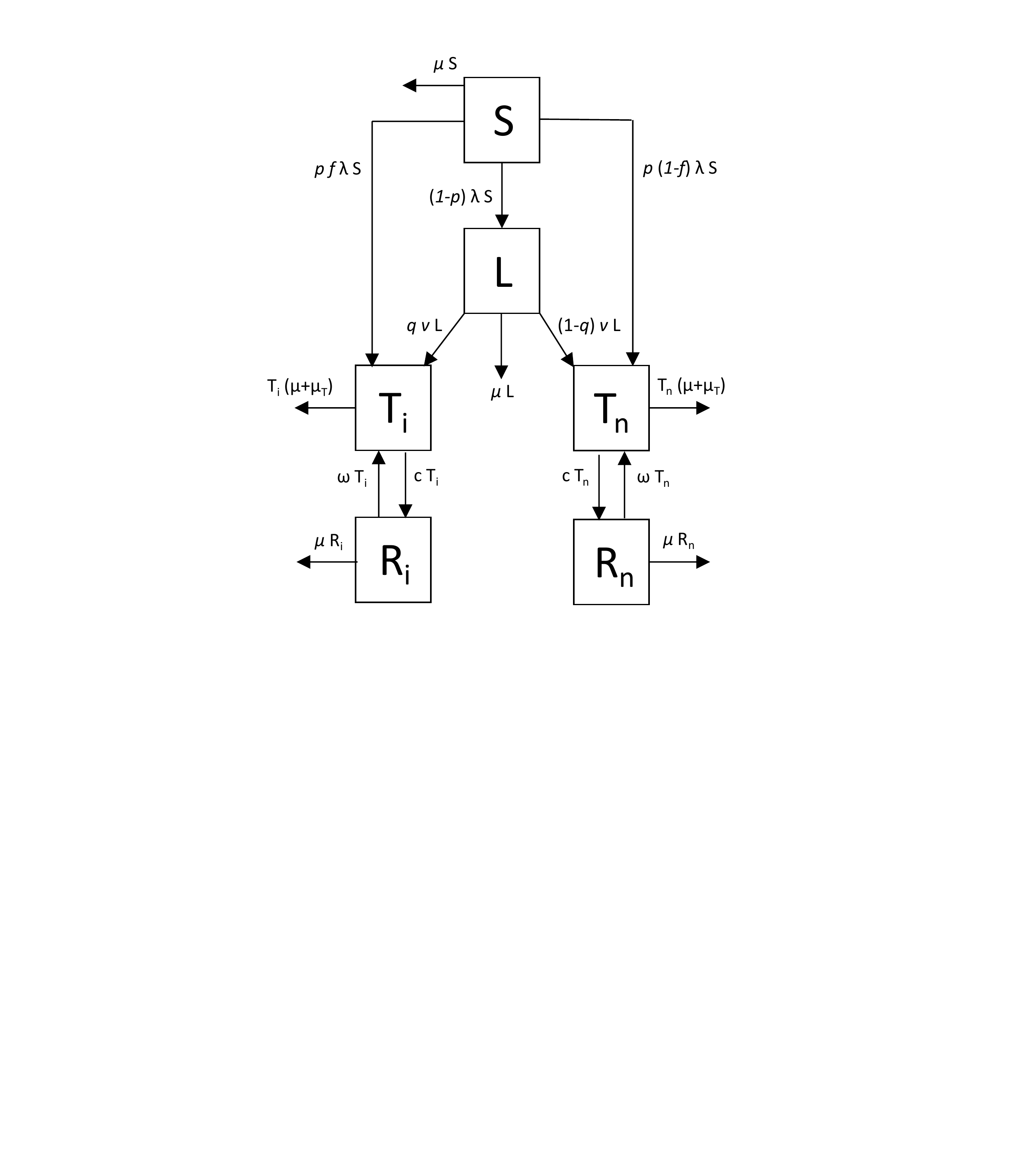}
\end{center}
\caption{Blower's model for Tuberculosis transmission \citep{blower1995intrinsic}. Only individuals in $T_i$ are infectious. This state communicates with state $R_i$ and individuals can alternate between states $T_i$ and $R_i$ until removal.}\label{fig:12}
\end{figure}

$$
\mathbf{U}= \bbordermatrix{ & S& L & T_i  & T_n &R_i & R_n \cr
S& 0 & \frac{\lambda  (1-p)}{\lambda +\mu } & \frac{(1-f) \lambda  p}{\lambda +\mu } & \frac{f \lambda  p}{\lambda +\mu } & 0 & 0 \cr
 L&0 & 0 & \frac{q v}{\mu +v} & \frac{(1-q) v}{\mu +v} & 0 & 0 \cr
 T_i &0 & 0 & 0 & 0 & \frac{c}{c+\mu +\mu _t} & 0 \cr
 T_n &0 & 0 & 0 & 0 & 0 & \frac{c}{c+\mu +\mu _t} \cr
 R_i&0 & 0 & \frac{\omega }{\mu +\omega } & 0 & 0 & 0 \cr
 R_n&0 & 0 & 0 & \frac{\omega }{\mu +\omega } & 0 & 0 \cr}
$$

The expression for the fundamental matrix $\mathbf{N}$ is cumbersome, but we only need the third column of the first row, containing the expected amount of visits to the only infectious state, $T_i$, starting in state S, which is:

\begin{equation}
\frac{\lambda  (\mu +\omega ) \left(c+\mu +\mu _t\right) ((1-f) \mu  p+p v (-f-q+1)+q v)}{(\lambda +\mu ) (\mu +v) \left(c \mu +(\mu +\omega ) \left(\mu +\mu _t\right)\right)}
\label{eqn:blower}
\end{equation}
\\
Observe the average time per visit to state $T_i$ is $(c+\mu +\mu _t)^{-1}$. Multiplying this by the average number of visits to $T_i$ yields the expected total time in the infectious state $T_i$:

$$
\frac{\lambda  (\mu +\omega ) ((1-f) \mu  p+p v (-f-q+1)+q v)}{(\lambda +\mu ) (\mu +v) \left(c \mu +(\mu +\omega ) \left(\mu +\mu _t\right)\right)}
$$
\\
\noindent
and to obtain $R_0$ we need to multiply the previous value by the contact rate in state $T_i$, which according to  \cite{blower1995intrinsic} is $\beta \Pi /\mu$. 
   
   \section{Discussion}
   
   We have presented a procedure that, given today's computational capabilities, has as its most complicated step the construction of the transition matrix $\mathbf{U}$. Most epidemic models start with a flux diagram between states with a differential equation for each arrow. In our approach, we substitute differential equations by probabilities. It is also very intuitive.
   
   The basic requirement for this approach is that the transition probabilities are constant, so that $\mathbf{U}$ does not contain the number of individuals in any stage, only parameters. There are models in which $\mathbf{U}$ is not constant and thus, this method can not be applied, for instance \cite{feng2000model} suggested an SEIT model in which individuals in state E can die at a constant rate $\mu$ or can move to state I at a rate $c I / N$, therefore, the probability of moving to state I is:
 
 $$
\frac{c I }{c I  + N \mu}.
$$
Therefore, we can not use a Markov chain approach to this model. This sort of models are not very common, and as it was shown in an earlier, less efficient version of the approach presented here \citep[see][]{hernandez2002markov}, $R_0$ in Feng's model is not maximized a the onset of the epidemic.

There is a correspondence between this approach and the Matrix Population Models theory for population growth \citep[see][]{caswell2009stage}, a stochastic version of deterministic modeling developed by \cite{leslie1945use,leslie1948some}. If \textit{infections} are considered \textit{births}, and every individual remains in a stage a unit of time per visit, then, we can build a matrix of fertility $\mathbf{F}$ containing the contact rate of state $j$ at position $(i,j)$ where state $i$ is the state were infected individuals come from (usually the susceptible class). Then:

$$
\mathbf{A}=\mathbf{FN'}
$$
is called the \textit{next generation operator}, since $\mathbf{x}(t+1) = \mathbf{A x}(t)$, where $\mathbf{x}(t)$ is the vector containing the number of individuals in each stage at generation $t$. Therefore, the dominant eigenvalue of $\mathbf{A}$ (the element in the upper left corner of $\mathbf{A}$) is the production of infections of an infected individual during its life time. Since we assumed that every visit to a state last a unit if time, we need to multiply this by the average time in the infectious stage, to obtain our $R_0$.

For instance, in the Tuberculosis model of Figure 13,  \citep{blower1995intrinsic}, our \textit{fertility} matrix would be:

$$
\mathbf{F}= \bbordermatrix{ & S& L & T_i  & T_n &R_i & R_n \cr
 S & 0 & 0 & \lambda  & 0 & 0 & 0 \cr
 L&0 & 0 & 0 & 0 & 0 & 0 \cr
 T_i&0 & 0 & 0 & 0 & 0 & 0 \cr
 T_n&0 & 0 & 0 & 0 & 0 & 0 \cr
 R_i&0 & 0 & 0 & 0 & 0 & 0 \cr
 R_n&0 & 0 & 0 & 0 & 0 & 0 \cr}
$$
After obtaining $\mathbf{N}$, the fundamental matrix of $\mathbf{U}$ for this model, the dominant eigenvalue of $\mathbf{A}=\mathbf{FN'}$ is identical to (\ref{eqn:blower}) times $\lambda$, as expected. Multiplying this by the average time per visit to state $T_i$, $(c + \mu + \mu_t)^{-1}$, yields the same $R_0$.

     \section{Appendix}
   
  \subsection{Appendix A1}
  If an individual produces $X$ descendants with probability $P_X$ during its life, then $P_e$, the probability of extinction of the population that starts with one individual, is:

$$
 P_e = (P_e | X=0) P_0 + (P_e | X=1) P_1 + (P_e | X=2) P_2 + \cdots
 $$
But by independence of fates of individuals, $ (P_e | X=k)= P_e^k$, thus,
\begin{eqnarray*}
 P_e& =& P_0 + P_e \ P_1 + P_e^2 \ P_2 + \cdots \\
 & = & \sum_{i=0}^\infty P_e^i \ P_i=M_X(P_e)
 \end{eqnarray*}
that is, the probability of extinction equates the \emph{Moment Generating Function} of $X$, the offspring production. Thus, the probability of extinction $P_e$ is the solution of $z = M_X(z)$. This system has always the solution $z=1$ and it may exist a second solution in $0 < z < 1$ . The probability of extinction is the minimum value of all the solutions in $[0,1]$. Branching process theory asserts that if the average offspring size is smaller than one, the probability of extinction is one since on average, at generation $1$ there will be $R_0$ individuals; at generation $2$, $R_0^2$; at generation $3$, $R_0^3$ and in general, at generation $n$, $R_0^n$, which tends to $0$ when $n \rightarrow \infty$ and $R_0 < 1$.  To prove that the mean offspring size is the relevant factor to determine wether extinction is certain or not, is straightforward: observe that we are looking or the solutions of $z$ such that $M_X(z)=z$ in the interval $0 \le z \le1$, and observe that in this interval, the second derivative of $M_X(z)$ is

$$
\frac{d^2}{dz} M_X(z)= \sum_{i=2} i (i-1)z^{i-2}P(X=i) >0
$$
that is, $M_X(z)$ is a concave function. Since $M_X(0)=P_0$ and $M_X(1)=1$, the shape of $M_X(z)$ is one of the dotted line depicted in Figure 13. Observe also that

$$
\frac{d}{dz} M_X(z) |_{z=1}= \sum_{i=2} i \ P(X=i) =E[X]
$$
that is, the slope of $M_X(z)$ at $z=1$ is the average offspring size, therefore, to intersect the $z$ line in another point in $0< z < 1$, it is necessary the slope at $z=1$ (average offspring size) be larger than $1$. 

  \begin{figure}[htbp]
\begin{center}
\includegraphics[width=10cm]{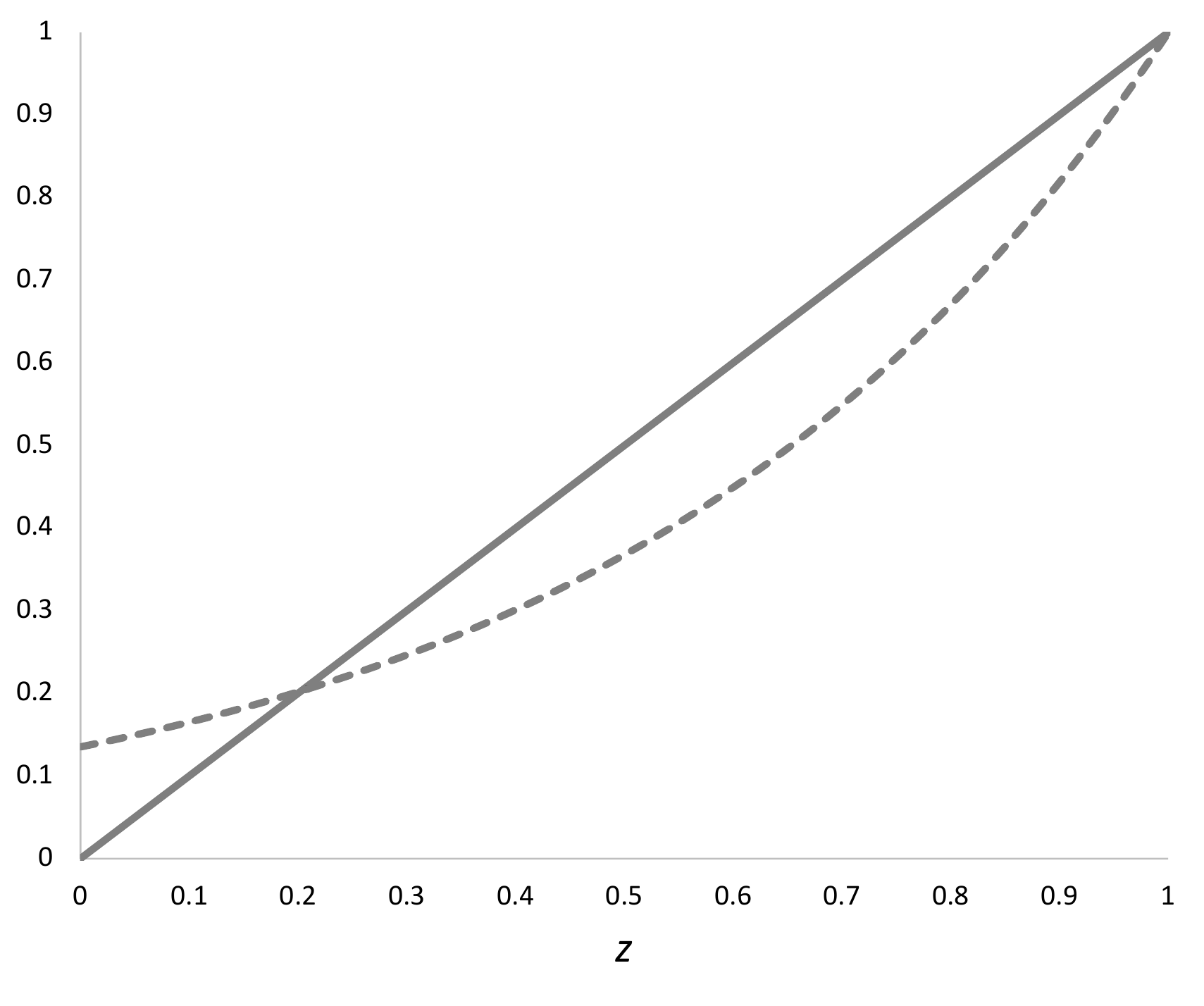}
\end{center}
\caption{Plot of the Moment Generating Function $M_X(z)$. We are interested in the minimum of the intersection points with $z$, that is, the probability of extinction. Since $M_X(z)$ is concave and goes always through $(0,0)$ and $(1,P_0)$, if the slope of $M_X(z)$ at $z=1$ is larger than one, there is always another solution of $M_X(z)=z$ in $(0,1)$. Since $M'_X(1)$ is by definition the expected value of $X$, then, if $E(X)>1$, the probability of extinction is less than $1$.}\label{fig:A1}
\end{figure}

 \subsection{Appendix A2}
 
 Let $(\mathbf{I}-\mathbf{U})\mathbf{x}=0$; that is $\mathbf{x}= \mathbf{Ux}$. Iterating this we can see that $\mathbf{x}= \mathbf{U}^n \mathbf{x}$. Since $\lim_{n \rightarrow \infty} \mathbf{U}^n \rightarrow \mathbf{0}$ we have $\mathbf{U}^n \mathbf{x} \rightarrow \mathbf{0}$, so $\mathbf{x} = \mathbf{0}$. Thus $(\mathbf{I}-\mathbf{U})^{-1}$ exists. 

The probability that the process will be in every state after $n$ steps is $\mathbf{U}^n$, thus, the expected number of visits to each state after $n$ transitions is:

$$
\mathbf{I}+\mathbf{U}+\mathbf{U}^2+\mathbf{U}^3+\ldots+\mathbf{U}^n
$$
observe that:
 $$
(\mathbf{I}-\mathbf{U})(\mathbf{I}+\mathbf{U}+\mathbf{U}^2+\mathbf{U}^3+\ldots+\mathbf{U}^n)=\mathbf{I}-\mathbf{U}^{n+1}
$$
multiplying both sides by $(\mathbf{I}-\mathbf{U})^{-1}$yields:
$$
\mathbf{I}+\mathbf{U}+\mathbf{U}^2+\mathbf{U}^3+\ldots+\mathbf{U}^n= (\mathbf{I}-\mathbf{U})^{-1} (\mathbf{I}-\mathbf{U}^{n+1})
$$
taking the limit when $n\rightarrow \infty$ yields:
$$
\sum_i ^\infty \mathbf{U}^i =  (\mathbf{I}-\mathbf{U})^{-1} \equiv \mathbf{N}
$$

\end{document}